\documentstyle[prb,multicol,aps]{revtex}

\def\sech{\mbox{sech}}

\begin{document}
\input{epsf.tex}
\epsfverbosetrue
\draft

\title{Interplay of disorder and nonlinearity 
in Klein-Gordon models: Immobile kinks}

\author{Serge F. Mingaleev and Yuri B. Gaididei}
\address{Bogolyubov Institute for Theoretical Physics, 
252143 Kiev, Ukraine}
\author{Eva Majern{\'\i}kov{\'a}}
\address{Department of Theoretical Physics of Palack{\'y} 
University, CZ-77207 Olomouc, Czech Republic}
\address{and Institute of Physics, SAS,
SK-84228 Bratislava, Slovak Republic}
\author{Serge Shpyrko}
\address{Institute for Nuclear Research, 
252028 Kiev, Ukraine}
\date{Received 30 April 1998; revised manuscript received 8 October 1998}

\maketitle

\begin{abstract}
We consider Klein-Gordon models with a 
$\delta$-correlated spatial disorder. We show that the properties 
of immobile kinks exhibit strong dependence on the 
assumptions as to their statistical distribution over the 
minima of the effective random potential. 
Namely, there exists a crossover from monotonically increasing 
(when a kink occupies the deepest potential well) to the non-monotonic
(at equiprobable distribution of kinks over the potential minima)
dependence of the average kink width as a function of the disorder
intensity. We show also that the same crossover may take place 
with changing size of the system. 
\end{abstract}

\begin{multicols}{2}
\narrowtext

\section{Introduction}
\label{sec:intro}

An extensive research work on the static and transport properties
of nonlinear excitations in various soliton-bearing
disordered systems  has been undertaken in the last  decade
(see Refs. \onlinecite{SV91} and \onlinecite{GK92}).
It is well known that when taken separately both
the nonlinearity and disorder contribute to the localization effects,
the character of localization being however essentially different. The
nonlinearity results in the possibility of existence of nonlinear
localized excitations usually referred to as solitons that are
rather robust and can propagate through the system undistortedly. At
the same time the disorder (in linear systems) evokes the Anderson
localization, which manifests itself in the behavior of the 
transmission coefficient of a plane wave decaying exponentially with 
the system width. These two localization mechanisms are competitive 
to some extent; taken together, the 
nonlinearity and disorder may lead to a number of qualitatively new 
effects, namely, the transmission coefficient tending to zero with 
increasing the system length according to a power law
\cite{DS86,DR87} rather than exponentially (which would be otherwise
the property of a linear system); there can arise a multistability
\cite{DS86,DR87,KPW91} in the wave transmission through a disordered 
slab; excitations in highly nonlinear or multidimensional nonlinear 
Schr{\"o}dinger (NLS) systems (which would either disperse or collapse 
otherwise) can be stabilized by disorder. \cite{BMM95,CGJRUV97,GHCR98}

In the present paper we study the static 
properties of a one-kink solution (or, equivalently, of a diluted kink
gas) of disordered Klein-Gordon models where the disorder is
assumed to be a $\delta$-correlated Gaussian spatial noise.
The disorder of the kind is akin, for example, to Josephson
junctions, where it is caused by the fluctuations of the gap between
two superconductor plates.
Quite recently Mints proposed \cite{Min98} a similar model with 
randomly alternating critical current density to account for a 
self-generated magnetic flux observed \cite{MHM+96} experimentally. 
We discuss his results in more detail in the Conclusion. 

In general, Klein-Gordon models have been repeatedly studied 
\cite{SV91,GK92,GKMSV92} for the disorder represented by a lattice of 
$\delta$-like impurity potentials with random positions of the 
impurities and either equal or randomly distributed intensities.  It 
was found out that in a number of cases the kink dynamics in 
disordered systems could be adequately described within the framework 
of the collective coordinates approach (Refs. 
\onlinecite{McLS78,PVBP89,KM89,SSBV92}, and references therein). This is the 
background of our restricting ourselves to the said approach in the 
scope of the present paper providing  however at each step the 
validation of analytical results comparing them to the numerical 
simulations of the original system. We emphasize that in opposite to, 
for example, Refs. \onlinecite{GK92} and \onlinecite{GKMSV92}, we use 
Rice's collective coordinates approach \cite{Rice83,BW90} with the 
kink width being the variational parameter. 

A similar approach has been recently applied for two other 
one-dimensional (1D) systems: Bussac {\it et al.} investigated 
\cite{BMM95} the effects of the polaron ground state in a deformable 
chain, while Christiansen {\it et al.} considered \cite{CGJRUV97} the 
stabilization of nonlinear excitations by disorder in the NLS 
model. It should be indicated that investigation of these two
(in fact closely related) systems leads to one and the same 
result: the width of stationary solitons decreases with 
growing intensity of the disorder. The importance of this 
conclusion resides in its prediction that the disorder can stabilize 
otherwise unstable solitons in 2D and 3D NLS models. Quite recently 
this prediction was borne out numerically 
\cite{GHCR98} for the 2D case. It must be emphasized that for NLS models 
the conclusion does not depend on the averaging procedure: one can 
equally perform averaging either on absolute ground states 
\cite{BMM95} or over all local minima of the effective random 
potential with equal weights. \cite{CGJRUV97,GHCR98} We show that it 
is not the case for the Klein-Gordon models: their properties exhibit 
strong dependence on the assumptions as to the statistical 
distribution of kinks over the minima of the effective random 
potential. For the purely dynamical problem these statistics are 
left beyond consideration and should be thus imposed as an additional 
assumption. We use Jaynes's maximum entropy inference \cite{Jaynes} for 
this purpose. 

The outline of the paper is the following. In Sec. 
\ref{sec:system} we present the model and derive the equations for 
the collective coordinates of the kink taking into account the 
disorder via the {\it effective random potential}. In Sec. 
\ref{sec:immobile} we investigate both analytically and numerically 
the case of immobile kinks and demonstrate the existence of the
crossover between monotonic and nonmonotonic dependence of the average 
kink width as a function of the disorder intensity. In Sec. 
\ref{sec:summary} we summarize the exposed results.

\section{Collective coordinates approach}
\label{sec:system}

We consider a Klein-Gordon (KG) model in the presence of space
disorder. The Hamiltonian of the system has the form
\begin{eqnarray}
\label{hamilt}
H = \int_{-\infty}^{\infty} dx \biggl\{ \frac{1}{2} ( 
\phi_t^2 + \phi_x^2 ) + [1-\eta \epsilon(x)] 
\Phi(\phi) \biggr\} \; ,
\end{eqnarray}
where the subscripts stand for partial derivatives with respect to 
the indicated variables and units are chosen so that the Hamiltonian 
is already in the scaled form. The potential $\Phi(\phi)$ has the 
form 
\begin{equation}
\label{Phi-SG}
\Phi(\phi)=1-\cos \phi 
\end{equation}
for the sine-Gordon (SG) model, and 
\begin{equation}
\label{Phi-phi4}
\Phi(\phi)=\frac{1}{4} (\phi^2-1)^2 
\end{equation}
for the $\phi^4$ model. We assume that $\epsilon (x)$ is 
delta-correlated spatial disorder 
\begin{equation}
\label{delta-correl}
\langle \epsilon(x) \epsilon(x') \rangle = \delta(x-x') \; ,
\end{equation}
(the brackets $\langle ... \rangle$ denote averaging over all
realizations of the disorder) with the Gaussian distribution 
\begin{equation}
p [\epsilon(x)]= \frac{1}{\sqrt{\pi}} 
\exp [-\epsilon^2(x)] \; .
\end{equation}
We have studied both SG (\ref{Phi-SG}) and $\phi^4$ 
(\ref{Phi-phi4}) models. Although properties of these two models show 
many similarities, they also exhibit a number of interesting 
distinctions related, in particular, to the existence of a breather 
state in the SG model and of Rice's internal mode \cite{Rice83,BW90} 
in the $\phi^4$ model. So, it was important to compare the effects of 
disorder in both these models. But since the qualitative features of 
the exposed models turned out to coincide in the scope of our present 
investigations, we dare not overload the paper with unnecessary 
repetitions and restrict ourselves with presenting in detail the 
sine-Gordon model only, keeping in mind although that every stage of 
the calculations applies for the $\phi^4$ model as well.

The SG system is governed by the equation of motion
\begin{equation}
\label{eq-mot}
\phi_{tt} - \phi_{xx} + [1- \eta \epsilon(x) ] 
\sin \phi + \gamma \phi_t = 0 \; ,
\end{equation}
where the damping term with the damping constant $\gamma$ has been
included.
It is well known that in the absence of disorder and damping 
($\eta =\gamma =0$) Eq. (\ref{eq-mot}) is completely 
integrable and possesses a topologically stable solution in the form 
of a kink given by 
\begin{equation} 
\label{sys:phi-k} 
\phi_K (x,t)= 4 \arctan \exp \biggl( \frac{x-X(t)}{L(t)} \biggr) \; , 
\end{equation} 
where $X(t)=X_0 + vt$ is the kink coordinate, $v$ is 
its velocity, and $L=\sqrt{1-v^2}$ is the kink width.

In the general case of Eq. (\ref{eq-mot}) for a
number of situations the kink emission is exponentially small, 
\cite{KM89} so that the kink dynamics can be
studied by the collective coordinate approach. In
the framework of this approach the variables $X(t)$ and $L(t)$ are 
understood as time-dependent variational parameters. 
Inserting Eq. (\ref{sys:phi-k}) into Hamiltonian (\ref{hamilt}) as a 
trial function, we obtain the effective Hamiltonian
\begin{equation}
H_{eff} = \frac{L}{16} p_X^2 + \frac{3L}{4 \pi^2} p_L^2
+ U(L) + V(\{ \epsilon \}, L, X) \; ,
\end{equation}
with  momenta
\begin{equation}
p_X= \frac{8}{L} \frac{d X}{dt} \; , \qquad
p_L= \frac{2 \pi^2}{3 L} \frac{d L}{dt} \; .
\end{equation}
Here
\begin{equation}
U(L)= \frac{4}{L} + 4 L \; 
\end{equation}
is the potential function in the case of no
disorder and
\begin{eqnarray}
\label{V:stoch}
V(\{ \epsilon \}, L, X) = -2 \eta \int_{-\infty}^{\infty}
dx \epsilon(x) \; \sech^2 \biggl( \frac{x-X}{L} \biggr) 
\end{eqnarray}
is the effective random potential arising because of the 
disorder term. Then, taking into account the damping, we arrive 
at the following equations of motion: \cite{Rice83}
\begin{equation}
\label{eq1-var}
\frac{d p_X}{dt} + \gamma p_X + \frac{d}{dX} V(\{ \epsilon
\}, L, X) = 0 \; ,
\end{equation}
\begin{eqnarray}
\label{eq2-var}
\frac{d p_L}{dt} &+& \gamma p_L + \frac{3 p_L^2}{4\pi^2} +
\frac{p_X^2}{16} \nonumber \\ &+& \frac{d}{dL} 
[ U(L)+ V(\{ \epsilon \}, L, X) ] = 0 \; .  
\end{eqnarray} 
In the following section we solve approximately these equations of 
motion for immobile kinks and compare the results to the results of 
direct numerical integration of Eq. (\ref{eq-mot}).

\section{Results for immobile kinks}
\label{sec:immobile}

As a consequence of the damping $\gamma$ the kink will eventually 
stop at some stable or metastable stationary position along the 
system. Here we do not consider this transient stage and 
assume that the kink is already immobile [$d p_L/dt = p_L 
= p_X = 0$]. In this case the equations of motion 
(\ref{eq1-var}) and (\ref{eq2-var}) take on the form 
\begin{equation}
\label{eq1-immob}
\frac{d}{dX} V(\{ \epsilon \}, L, X) = 0 \; ,
\end{equation}
\begin{equation}
\label{eq2-immob}
\frac{d}{dL} [ U(L)+ V(\{ \epsilon \}, L, X) ] = 0 \; .  
\end{equation} 
Considering the center-of-mass motion described by Eq. 
(\ref{eq1-var}) we observe that for each realization of the random 
potential $\epsilon(x)$ the stable stationary position $X=X_m(\{ 
\epsilon \}, L)$ of the kink is defined by the point where $V(\{ 
\epsilon \}, L, X)$ has a minimum with respect to $X$. Thus we can 
now insert the value $X=X_m(\{ \epsilon\}, L)$ into Eq. 
(\ref{eq2-immob}) and, solving the resulting equation
\begin{equation}
\label{immob:eq-ell}
L = \biggl( 1 + \frac{1}{4} \frac{d}{dL} V(\{ \epsilon \}, L, X_m) 
\biggr)^{-1/2} \; , 
\end{equation}
get the value of the stationary kink width $L(\{\epsilon \})$ at 
given position of the kink. Encountering a similar problem for 
the case of NLS system, Christiansen {\it et al.} 
invoked \cite{CGJRUV97} the mean-field approximation 
\begin{equation} 
\label{approx}
\langle \frac{d}{dL} V(\{ \epsilon \}, L, X_m) \rangle \approx 
\frac{d}{dL} \langle V(\{ \epsilon \}, L, X_m) \rangle \; . 
\end{equation} 
Further the estimation of the quantity $\langle V \rangle$ was
performed using Rice's averaging theorem. \cite{Ric94,KS86} 
However, being rather good for the NLS model \cite{CGJRUV97} the 
mean-field approximation (\ref{approx}) fails for the KG models. Thereby we
were forced to use a more precise averaging procedure calculating 
$\langle dV/dL \rangle$ directly. 

Expanding Eq. (\ref{immob:eq-ell}) into  series up to the second 
order in $\eta$, 
\begin{equation}
L\approx 1-\frac{1}{8}\frac{dV}{dL} + \frac{3}{128}
\biggl( \frac{dV}{dL}\biggr)^2 \; , 
\end{equation}
and solving by iterations we get after averaging [which by means of 
Eq. (\ref{delta-correl}) can be performed for the terms containing 
$\eta^2$ exactly] the average kink width
\begin{equation}
\label{immob:ell}
L_{var} \equiv \langle L(\{ \epsilon \}) \rangle \approx 
1 + \frac{\eta}{2} \langle \lambda(\{ \epsilon \}, X_m) 
\rangle +\frac{\pi^2}{180} \, \eta^2 \; ,
\end{equation}
where the averaging of the function
\begin{equation}
\lambda(\{ \epsilon \}, X)= \int_{-\infty}^{\infty} 
dx \epsilon(x) (x-X) \frac{\sinh(x-X)}{\cosh^3 (x-X)} \; 
\end{equation}
is performed over all realizations of the disorder in the points 
$X=X_m$ in which the potential
\begin{equation}
\label{immob:mu1}
V(\{ \epsilon \}, L, X) \approx 2 \eta 
\mu(\{ \epsilon \}, X) \; 
\end{equation}
with 
\begin{equation}
\label{immob:mu2}
\mu(\{ \epsilon \}, X)= - \int_{-\infty}^{\infty} dx 
\epsilon(x) \sech^2(x-X) \;  
\end{equation}
takes on its minima on $X$.

\begin{figure}
\setlength{\epsfxsize}{80mm}
\centerline{\epsfbox{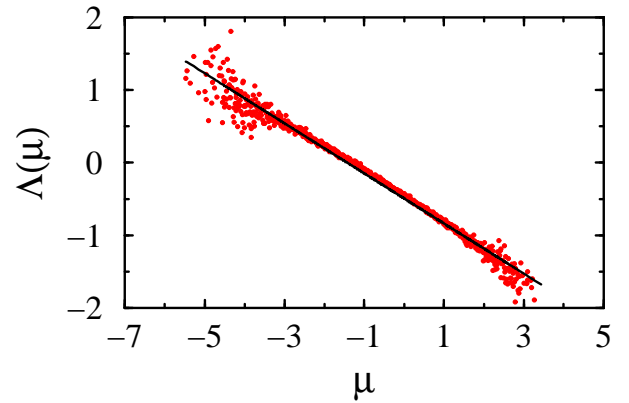}}
\caption{
The dependence $\Lambda(\tilde{\mu})$ found numerically. 
The calculations are performed for a million minima 
of the function $\mu(\{ \epsilon \}, X)$.} 
\label{fig:lambda}
\end{figure}

Thus we arrive at the problem of performing the average of $\lambda 
(\{ \epsilon \}, X_m)$ over the minima $X_m$ of the function $\mu 
(\{ \epsilon \}, X)$. It is convenient for later use to perform 
this averaging in two steps calculating at the outset
\begin{equation}
\label{immob:lambda-mu}
\Lambda (\tilde{\mu}) = \int_{-\infty}^{\infty} 
\tilde{\lambda} \, P_{l} (\tilde{\lambda} \mid \tilde{\mu}) 
d \tilde{\lambda} \; 
\end{equation}
and thereafter 
\begin{equation}
\label{immob:lambda-x}
\langle \lambda (\{ \epsilon \}, X_m) \rangle = 
\int_{-\infty}^{\infty} \Lambda (\tilde{\mu}) 
P_m (\tilde{\mu}) d \tilde{\mu} \; .
\end{equation}
Here $P_{l} (\tilde{\lambda} \mid \tilde{\mu})$ is the conditional 
probability that $\lambda (\{ \epsilon \}, X_m)$ has the 
value $\tilde{\lambda}$ if $\mu (\{ \epsilon \}, X_m)$ 
equals to $\tilde{\mu}$. Correspondingly, $\Lambda (\tilde{\mu})$ is 
the value of $\lambda (\{ \epsilon \}, X_m)$ averaged over all 
realizations of the disorder for which $\mu 
(\{ \epsilon \}, X_m)$ is equal to $\tilde{\mu}$. It is difficult to 
calculate $\Lambda (\tilde{\mu})$ analytically but numerical 
simulations show (see Fig. \ref{fig:lambda}) that up to very good 
accuracy the dependence $\Lambda (\tilde{\mu})$ is linear: 
\begin{equation}
\Lambda (\tilde{\mu}) \simeq - 0.344 \, \tilde{\mu} - 0.500 \; . 
\end{equation}

\vspace{-2mm}

Substituting it into Eq. (\ref{immob:lambda-x}) we obtain that
\begin{equation}
\label{immob:lambda-x2}
\langle \lambda (\{ \epsilon \}, X_m) \rangle \simeq - 0.344 \,
\langle \mu (\{ \epsilon \}, X_m) \rangle - 0.500 \; ,
\end{equation}
where 
\begin{equation}
\label{mu-average}
\langle \mu (\{ \epsilon \}, X_m) \rangle = 
\int_{-\infty}^{\infty} \tilde{\mu} \, 
P_m (\tilde{\mu}) d \tilde{\mu} \; 
\end{equation}
is the average value of the function $\mu (\{ \epsilon \}, X)$ over 
its minima $X_m$. Here the probability density $P_m (\tilde{\mu})$ 
is a product of two factors. The first 
one is the probability density that some arbitrary chosen minimum 
$X_m$ of the function $\mu(\{ \epsilon \}, X)$ will be equal to 
$\tilde{\mu}$. We denote this probability density as 
$p_{min}(\tilde{\mu})$. The second factor is the conditional probability 
that if the minimum is equal to $\tilde{\mu}$ it will be actually 
occupied by the kink. It is evident that in a real system 
the kink is more likely to occupy the deeper minimum than the shallow one.  
So to be consequential one must ascribe to every minimum of the function 
$\mu(\{ \epsilon \}, X)$ some probability weight and average taking 
into account those probabilities. But the values of these 
probabilities are in general determined by the whole prehistory of 
the kink. These values are not contained in the dynamical equations 
of motion that state only that the kink should take on some minimum 
regardless to its depth. It would be a cumbersome problem to 
calculate them appropriately. That is why two limit cases are in 
general considered: either the kink seats itself into the deepest 
well \cite{BMM95} or it rather occupies any of them with 
equal probabilities. \cite{CGJRUV97} As it was already 
remarked in the Introduction, both limit cases lead to qualitatively 
the same results for the NLS model. 

It is not the case of KG models where, as it will be shown later, 
different assumptions as to the {\it a priori} weights of minima lead 
to qualitatively different behavior of the kink width. 
Since we are merely lacking information sufficient enough to 
reconstruct these weights in an objectivistic fashion, a remedy would 
be Jaynes's maximum entropy inference, \cite{Jaynes} according to 
which the simplest self-consistent unbiased choice is to assume that 
the kink will occupy a potential well corresponding to the minimum 
$X_m$ of the function  $\mu(\{ \epsilon \}, X)$ [for {\it given} 
profile $\epsilon(x)$] with probability proportional to $e^{- \beta 
\mu(\{ \epsilon \}, X_m)}$ thus introducing an additional 
parameter $\beta$ (following Jaynes we shall call it a conjugate 
parameter). By this expedience we present 
some natural interpolation covering two mentioned limit cases: of 
equiprobable distribution ($\beta =0$) and of averaging over the 
deepest minima only ($\beta \to \infty$). So we can write
\begin{equation}
P_m(\tilde{\mu}) 
= \frac{1}{{\cal Z}} e^{-\beta \tilde{\mu}} p_{min} 
(\tilde{\mu}) \; ,  
\end{equation}
where 
\begin{eqnarray}
{\cal Z}(\beta)  = \int_{-\infty}^{\infty} d \tilde{\mu} 
e^{-\beta \tilde{\mu}} p_{min} (\tilde{\mu}) \;  
\end{eqnarray}
plays part of the partition function. 

To calculate $p_{min} (\tilde{\mu})$ we follow Ref. 
\onlinecite{CGJRUV97} and 
make use of the Rice's averaging theorem \cite{Ric94,KS86} [valid for 
the case, well attested by the numerics, of $\mu(\{ \epsilon \}, X)$ 
being a stationary centered Gaussian process] stating that the 
probability density of some given minimum of the function 
$\mu(\{ \epsilon \}, X)$ to be equal to $\tilde{\mu}$ is
\begin{equation}
p_{min} (\tilde{\mu}) = \frac{1}{\sqrt{2 \pi M_0}} \, \sigma \biggl(
\frac{\tilde{\mu}}{\sqrt{2 M_0}} \, , \, \sqrt{1- \frac{M_2^2}{M_0
M_4}} \biggr) \; ,
\end{equation}
where the function
\begin{eqnarray}
\sigma(y, \kappa) &=& \kappa e^{- (y/\kappa)^2} 
\nonumber \\ &-& 2 \sqrt{1- \kappa^2}
y e^{-y^2} \int_{ (y/ \kappa) \sqrt{1-
\kappa^2}}^{\infty} e^{-t^2} dt \; ,
\end{eqnarray}
and the spectral momenta
\begin{eqnarray}
\label{M0-M4}
M_0 = \langle [\mu(\{ \epsilon \}, X) ]^2 
\rangle = \frac{4}{3} \; , \nonumber \\ 
M_2 = \langle [\mu_X (\{ \epsilon \}, X) ]^2 \rangle = 
\frac{16}{15} \; , \\
M_4 = \langle [\mu_{XX} (\{ \epsilon \}, X) ]^2 \rangle = 
\frac{64}{21} \; , \nonumber
\end{eqnarray}
were introduced.

Hence the partition function takes on the form
\begin{equation}
{\cal Z}(\beta)=\frac{1}{\sqrt{\pi}} 
\int_{-\infty}^{\infty} dx e^{-\sqrt{8/3} \, \beta x}
\sigma \left(x,\frac{3}{5}\sqrt{2} \right)
\end{equation}
and can be expanded into series in $\beta$ 
yielding the average value
\begin{eqnarray} 
\langle \mu(\{ \epsilon \}, X_m) \rangle = 
- \frac{d}{d \beta} \ln {\cal Z} (\beta) \\ \nonumber
= - \frac{1}{5} \sqrt{\frac{7 \pi}{6}} - 
\frac{2}{75} (64-7 \pi) \, \beta 
+ {\cal O} (\beta^2) .\;  
\end{eqnarray} 

And now, substituting it into Eqs. (\ref{immob:lambda-x2}) and 
(\ref{immob:ell}) we obtain for the average kink width
\begin{eqnarray} 
\label{immob:ell-var}
L_{var} \approx 1 + (0.193 \beta - 0.059) \, \eta + \frac{\pi^2}{180} 
\, \eta^2 \; .  
\end{eqnarray} 
Thus it is seen that there is a qualitative change of the kink width 
behavior as function of the disorder intensity $\eta$ 
according to whether the value of conjugate parameter $\beta$ is 
below or above some critical value $\beta_{cr}\approx 0.3$. 

\begin{figure}
\setlength{\epsfxsize}{80mm}
\centerline{\epsfbox{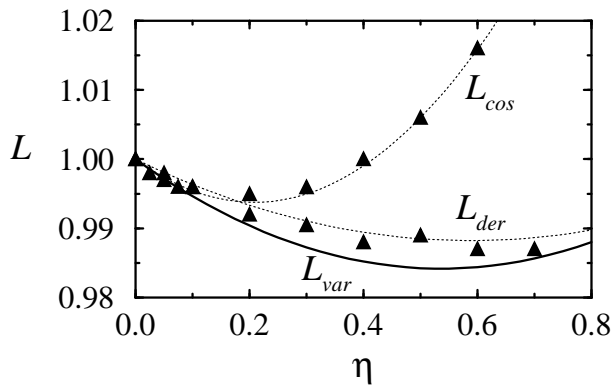}}
\caption{
The dependence of the average (on 1000 realizations of 
the disorder) kink width $\langle L(\{ \epsilon \}) 
\rangle$ vs disorder intensity $\eta$ under the assumption that the 
kink occupies every potential well created by disorder with equal 
probability. The system length $R=15$.} 
\label{fig:ell-loc-min}
\end{figure}

At small $\beta$ the 
average kink width is a nonmonotonic function of the disorder 
intensity $\eta$: it decreases at small intensities but starts to
increase thereafter. 
This result is well attested by the direct numerical 
calculations of stationary kink solutions of the initial equation 
of motion (\ref{eq-mot}). In Fig. \ref{fig:ell-loc-min} we compare the 
analytical prediction given by Eq. (\ref{immob:ell-var}) with the
numerical results for the case of equiprobable distribution of the
kinks over the potential wells ($\beta=0$). The numerical results have 
been obtained as an average of 1000 realizations of the disorder. 
Two different expressions for the kink width were calculated: 
\begin{equation}
\label{el-2}
L_{cos}= \frac{1}{4} \int_{-\infty}^{\infty} dx \{ 
1-\cos[\phi(x)] \} \; ,
\end{equation}
and
\begin{equation}
\label{el-der}
L_{der}= 8 \biggl\{ \int_{-\infty}^{\infty} dx \biggl[ 
\frac{d \phi(x)}{dx} \biggr]^2 \biggr\}^{-1} \; .
\end{equation}
From the point of view of the collective coordinate approach
$L_{cos}$ and $L_{der}$ should coincide with the value of
$L$ introduced in Eq. (\ref{sys:phi-k}). Indeed, as it is seen from the 
figures, it does take
place for $0 \leq \eta \lesssim 0.2$; thus, these are limits where the
collective coordinate approach works well.

\vspace{-2mm}
\vspace{-2mm}
\begin{figure}
\setlength{\epsfxsize}{80mm}
\centerline{\epsfbox{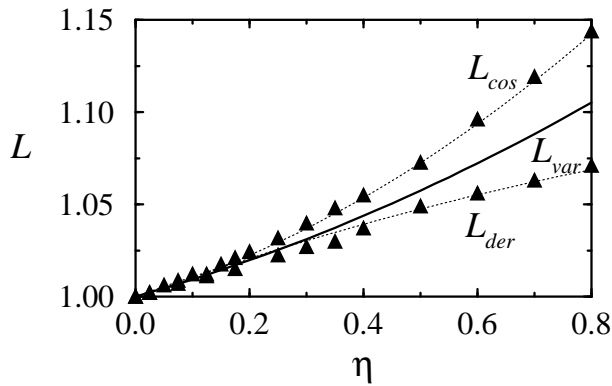}}
\caption{
The dependence of the average (on 1000 realizations of 
the disorder) kink width $\langle L(\{ \epsilon \}) 
\rangle$ vs disorder intensity $\eta$ under the assumption that the 
kink sits into the deepest potential well created by disorder. 
The system length $R=15$. 
} 
\label{fig:ell-abs-min-big} 
\end{figure}

Going on with the analysis of Eq. (\ref{immob:ell-var}) we see that 
at big values of the conjugate parameter ($\beta>\beta_{cr}$) when 
the kink rather occupies deep potential wells, the kink width should 
grow monotonically with the disorder. But it is evident that in this 
case the analytical approach discussed above is applicable 
to systems of infinite length only. For 
the finite system the case of big $\beta$ represents the situation
when the kink sits in the deepest potential well. Obviously its
average depth essentially depends on the system length. We can
make an estimation of this dependence 
drawing on the formula \cite{KS86} for the average 
number of minima of the function $\mu(\{ \epsilon \}, X)$ 
on an interval $R$ whose  values lie below some $\tilde{\mu}$:
\begin{equation}
N_{min}= \frac{R}{2 \pi}
\sqrt{\frac{M_2}{M_0}} e^{- \tilde{\mu}^2 /(2 M_0)} \; .
\end{equation}
Inserting there $N_{min}=2$ ($\tilde{\mu}$ is an absolute minimum 
on the interval $R$ but not on the longer interval) 
one can estimate its average value 
\begin{eqnarray}
\langle \mu(\{ \epsilon \}, X_{abs.m}) \rangle &\simeq& - 
\biggl( 2 M_0 \ln \biggl\{ \frac{R}{4 \pi} \sqrt{\frac{M_2}{M_0}} 
\biggr\} \biggr)^{1/2} \nonumber \\ 
&=& - \biggl( \frac{8}{3} \ln \biggl\{ 
\frac{R}{2 \sqrt{5} \pi} \biggr\} \biggr)^{1/2} \;
\end{eqnarray}
and, substituting it into Eqs. (\ref{immob:ell}) and
(\ref{immob:lambda-x2}) one can find that the average kink width 
equals
\begin{equation}
\label{l:abs:dim}
L_{var}\simeq 1 + \biggl( 0.28 \, \ln^{1/2} \biggl\{ 
\frac{R}{2 \sqrt{5} \pi} \biggr\} -0.25 \biggr) \eta + 
\frac{\pi^2}{180} \, \eta^2 \; .
\end{equation}

\vspace{-2mm}
\begin{figure}
\setlength{\epsfxsize}{80mm}
\centerline{\epsfbox{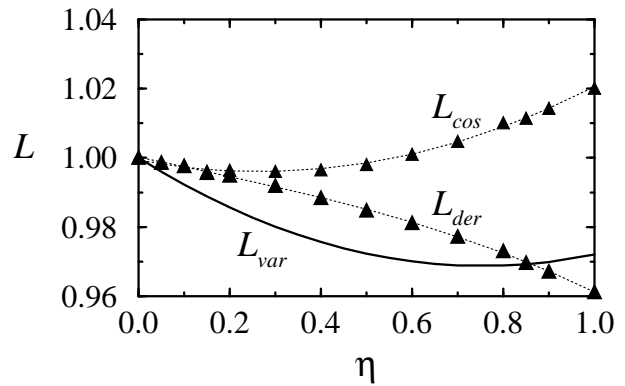}}
\caption{
The dependence of the average (on 1000 realizations of 
the disorder) kink width $\langle L(\{ \epsilon \}) 
\rangle$ vs disorder intensity $\eta$ under the assumption that the 
kink sits into the deepest potential well created by disorder. 
The system length $R=5$. 
} 
\label{fig:ell-abs-min-small} 
\end{figure}

It is seen that for finite-size systems the character of dependence 
$L_{var}(\eta)$ depends on the size of the system $R$. It is 
interesting that even for the case of averaging over the absolute minima 
considered here, the function $L_{var}(\eta)$ grows monotonically with 
$\eta$ only for the systems that are large enough ($R \gtrsim 7.5$). 
The reason is that for a small system the number of potential wells 
of the effective random potential is too small to yield the average 
over absolute minimum that would be essentially smaller than the 
average value calculated over all minima. Indeed, Figs. 
\ref{fig:ell-abs-min-big} and \ref{fig:ell-abs-min-small}, in which 
we compare Eq. (\ref{l:abs:dim}) to the results of the numerical 
calculations for $R=15$ and $R=5$, lend support to the validity 
of the approach leading to Eq. (\ref{l:abs:dim}). 
One can see from these figures that the average kink 
width grows monotonically with $\eta$ for $R=15$ but is nonmonotonic 
(similar to the case depicted on Fig. \ref{fig:ell-loc-min}) 
for small $R=5$. But in this latter case the boundary conditions
become very important and most likely they are responsible for 
the difference between Figs. \ref{fig:ell-loc-min} and 
\ref{fig:ell-abs-min-small}.

\section{Conclusion}
\label{sec:summary}

In the paper we consider the Klein-Gordon models with the 
$\delta$-correlated spatial disorder and investigate both 
analytically and numerically the width of immobile kinks as a 
function of the intensity of disorder. 
The analytical collective coordinates approach is based on 
Rice's averaging theorem from the theory of random processes 
\cite{Ric94,KS86} as well as on the maximum entropy 
inference proposed by Jaynes. \cite{Jaynes} 

We have shown that the properties 
of the kinks exhibit strong dependence on the 
assumptions as to their statistical distribution over the 
minima of the effective random potential. 
Namely, there exists a crossover from monotonically increasing 
(when a kink occupies the deepest potential well) to the nonmonotonic
(at equiprobable distribution of kinks over the potential minima)
dependence of the average kink width as a function of the disorder
intensity. We have shown also that the same crossover may take place 
with the changing size of the system: the average kink width monotonically 
increases for the systems of big size but is nonmonotonic for the 
small ones. 

It is interesting to compare the effects of the disorder in the KG model 
with the effects in the nonlinear Schr{\"o}dinger (NLS) model. As it was 
recently shown in Refs. \onlinecite{BMM95,CGJRUV97,GHCR98}, the 
$\delta$-correlated 
spatial disorder in the NLS systems creates an additional factor 
contributing to the decrease of the excitation width. This effect, 
being insensitive to the manner of the statistical distribution of 
kinks over the minima of effective random potential, favors the 
stabilization of excitations in highly nonlinear or multidimensional 
systems, which would either disperse or collapse otherwise.
The stabilizing function of disorder is of no doubt important for 
practical applications and to elucidate the extent to which it is 
universal seems to be an intriguing question.
The considered example of the KG systems demonstrates that there 
exists a class of systems, for which, in contrast to the NLS system, the 
effects of disorder can lead in different cases to diametrically 
opposed behavior.

In the case of the SG model we can consider the term $\eta \epsilon(x)$ 
as a change of the Josephson current density due to fluctuations of the 
thickness of an insulating layer. 
Quite recently Mints studied \cite{Min98} such a model to account for a 
self-generated magnetic flux observed \cite{MHM+96} by Mannhart 
{\it et al}. He has shown that in the case of  $\eta<1$ a state with
a self-generated flux exists and can be studied experimentally {\it in
the presence of Josephson vortices}. However, as is shown in the
present paper, the Josephson energy [equal to $4 L_{cos}$ in Eq. 
(\ref{el-2})] and the magnetic energy [equal to $4/L_{der}$ in Eq. 
(\ref{el-der})] of the Josephson vortices are functions of the
intensity of fluctuations of the insulating layer thickness. 
And their contribution into the experimentally observable 
magnetic flux will strongly (up to the change of a sign) depend on 
the statistical distribution of the vortices along the 
Josephson junction. Thus, for the proper description of the problem
one must develop a thermodynamic model.

\section*{Acknowledgments}

We (S.M., Yu.G., and S.Sh.) thank the Department of Theoretical 
Physics of the Palack\'y University in Olomouc for the 
hospitality. S.M. and Yu.G. acknowledge support from the Fund 
for Development of the M\v SMT \v CR No.~155/1997 and from the 
Ukrainian Fundamental Research Fund (Grant No.~2.4/355). 
E.M. and S.Sh. acknowledge support from Grant No.~202/97/0166 
of the GA\v CR. Partial support from Grant No.~2/4109/97 
of the VEGA Grant Agency is also acknowledged.



\end{multicols}
\end{document}